\documentstyle[prl,aps]{revtex}

\begin{document}

\draft

\title{Response to Leahy's Comment on the Data's Indication of
Cosmological Birefringence}

\author{Borge Nodland}

\address{Department of Physics and Astronomy, and Rochester Theory
Center for Optical Science and Engineering, University of Rochester,
Rochester, NY 14627}

\author{John P. Ralston}

\address{Department of Physics and Astronomy, and Kansas Institute for
Theoretical and Computational Science, University of Kansas, Lawrence,
KS 66044}

\date{Submitted to Physical Review Letters (1997)}

\maketitle

\begin{abstract}
Recently, J. P. Leahy commented (astro-ph/9704285) on the indication of
anisotropy in the propagation of radio waves over cosmological
distances, published by Nodland and Ralston (B. Nodland and J. P.
Ralston, {\it Phys. Rev. Lett.} {\bf 78,} 3043 (1997);
astro-ph/9704196). Unfortunately, Leahy ignores the need for a
statistical comparison, egregiously selects data, compares incompatible
measures, uses a variable incapable of resolving birefringence, and
makes mistakes in the calculations he presents.
\end{abstract}              

\pacs{PACS numbers: 98.80.Es, 41.20.Jb}

Leahy's Comment \cite{lea} is based largely on a misunderstanding and
misuse of the variable $\beta$. We defined $\beta$ as a signed measure
of polarization rotation in our paper \cite{nod}. Yet, many authors,
including Leahy \cite{lea}, Eisenstein and Bunn \cite{eis}, Carroll and
Field \cite{car}, and Wardle et al. \cite{war} disregard the
definition, and substitute something else -- most commonly
$\chi-\psi$, while still calling their variable ``$\beta$.'' It is
interesting that the traditional notation for $\chi-\psi$ was
``$\Delta$.''  We invented the symbol ``$\beta$'' just to separate our
signed variable from ``$\Delta$.'' The practice of using any old
variable, and calling it by the new name ``$\beta$'' is puzzling, and, we
think, rather misleading for readers.

Since we did not use $\chi-\psi$ (or $\chi-\phi$ created by Leahy,
where $\phi$ is far from similar to $\psi$), the Comment by Leahy and
the above mentioned critical papers do not address our correlation, but
examine other things. In some cases, people have used variables
incapable of revealing a birefringent correlation even if one worked
with perfectly correlated anisotropic data.

Why does it matter? The underlying difficulty is a simple but subtle 
point of mathematics, involving the projective nature of angles 
between two planes. The issue has often been overlooked by scientists 
trained in the mathematics of angles between vectors.

A concrete illustration uses laboratory data on optically birefringent
material, say slabs of quartz. Suppose there is a number of identical
slabs one can put in front of a polarizer. Let us record angles as
increasing in the counterclockwise sense, facing the ray, and running
from zero to 360 degrees for a full circle. Let a paint spot on the
initial polarizer be located at a reference angle called ``$i$.''
Rotating the initial polarizer to $i+180^\circ$ (not $i+360^\circ$), the
physical set--up is unchanged. This reveals the projective character of
polarization angles: such angles equal themselves after adding an
integer number of $\pi$ (not $2 \pi$) radians.

Inserting one slab, suppose one finds with an analyzing polarizer that
the light is bright at angle $f = i+22^\circ$; again, as measured in a
counterclockwise direction facing the ray. Now, if we take the
difference of $i$ and $f$, we record $22^\circ$; if we take the
difference of $f$ and $i$, it is $-22^\circ$. Immediately, one could
also ask if the electric field of the polarized incoming light rotated
clockwise, that is by $-158^\circ$. Taking the difference of $i$ and
$f$ the other way, it would be $158^\circ$. Indeed, with a one-slab
experiment, nobody can tell. We observe, then, that from a raw data
point alone, one cannot discriminate between different assigned
rotation angles given by the values $\beta = \pm (i - f) \pm \pi$.

With a series of thinner and thinner slabs, experiments could resolve
the magnitude and sense of rotation. This is a very important and 
non--trivial point! We continue, supposing it is known that one slab gives
$\beta= +22^\circ \pm 2^\circ$. Adding more and more slabs might give net
rotations of $\beta = +22^\circ \pm 2^\circ$, $+44^\circ \pm 3^\circ$, $\cdots$,
$+88^\circ \pm 5^\circ$, $+110^\circ \pm 7^\circ$, $+132^\circ \pm 11^\circ$,
$+154^\circ \pm 15^\circ$. With such data in the laboratory notebook, we
find a nice straight line plot of $\beta$ versus slab number. In this
way, one uses the information about the data inductively, to assign the
rotation angle properly and according to the hypothesis, without going
back to zero rotation angle for each new slab. This is much like our
procedure.

However, proper data recording will not happen automatically: if an
uninformed person were invited into the lab, he or she might perfectly
well record $-92^\circ$ instead of $+88^\circ$, $-70^\circ$ instead of
$+110^\circ$, etc., as the polarizer was easier to turn that way, or
even write down $92^\circ$, $70^\circ$, $\cdots$, (taking the absolute value)
in a misguided effort to eliminate ``unphysical minus signs.'' (Never
mind the student who records all variables as negative simply by
habit.)  And, while the uninitiated might think this is far--fetched,
it exactly represents the history of the subject in the literature, in
which every stripe of difference between angle variables, including
truncating to positive absolute values, or forcing all variables to be
acute by definition, have occurred.

If one is trying to detect statistically a straight line dependence of
rotation versus slabs, using a table of numbers written down for the
angles $i$ and $f$, then carelessness in mixing definitions for the
rotation will be catastrophic. Thus Carroll and Field \cite{car} (cited
by Leahy in \cite{lea}) find that $\chi-\psi$ does not give a signal:
but this is not unexpected, as the sign and value of their variable is
absolutely meaningless, representing as it does the conventions of
numbers written down in the astronomers' tables.

Table 1, supplied by Leahy, illustrates the same errors. In this new
variant of ``$\beta$,'' called ``$\beta(\text{obs})$,'' Leahy compares
the observed polarization orientation relative to an observed intensity
gradient angle $\phi$. But our variable $\beta$ used an observed galaxy
axis angle $\psi$ -- how is one supposed to relate one variable to the
other? No explanation is given, and we find it very misleading to call
this a test of our work. Very unfortunately, the new variable
$\beta(\text{obs})$ used does not meaningfully account for any
clockwise/anticlockwise rotation. One finds that all angles
$\beta(\text{obs})$ have been selected to give the minimum value.
Leahy's $\beta(\text{obs})$ is an angle guaranteed not to allow a test
for birefringence.

If this was not enough, we find that our values of $\beta$ have not
even been reported faithfully. In his Table 1, Leahy lists values
``$\beta(\text{pred})$'' for a set of galaxies, which he states are
``the predicted rotation according to the fit of [1]'' (where [1] is
his reference to our paper \cite{nod}). The ``predicted'' $\beta$
values from our paper would seem to be the values $\beta_{\text{NR}}=
(1/2) (1/\Lambda_s) r \cos(\gamma)$ (see our Eq. 1 in \cite{nod}). We
recalculated the numbers:  in 7 out of 7 cases they did not at all
agree. Take, eg. 3C47. Leahy lists a positive $\beta(\text{pred})$ for
this galaxy ($+49^\circ$), but 3C47's direction is such that
$\cos(\gamma)$ is negative, and thus $\beta(\text{pred})$ should be
negative. Simply put, Leahy lists a set of numbers
``$\beta(\text{pred})$,'' which are calculated wrongly, and we wonder
why.

We have other problems with Leahy's Table 1, because the numbers Leahy
used for $\beta(\text{obs})$ are not, in fact, available from the
references.  Besides using data that cannot be verified, we cannot
accept the whole methodology of selecting a few special parts of a few
special objects to confront a statistical correlation such as ours.
That's right: the relation between the electric field and the
gradients in local intensity that is cited is entirely due to selection
of parts of objects that look pretty. One only gets a hint of it in
Leahy's Fig. 1, which suppresses the bulk of the radio map of the
object. Where the appearance is not pretty, the data is declared no
good.

Leahy's method is not close to confronting what we observed. The
histogram in Leahy's Fig. 1b for 3C47 is, according to Leahy,
``consistent with zero cosmic birefringence.'' But anisotropic
birefringence involves comparing $\beta$'s (not $\chi-\phi$'s) for {\it
many directions}. How can one say anything about anisotropic
birefringence -- which means inequivalence of different directions --
when one looks in only one direction, toward 3C47?

Leahy supplies a single intensity map for the south lobe of quasar 
3C47. The data is lovely: what a pretty pattern. As anyone who has 
ever seen VLBI data can assure, this is an  example of data selection 
of the highest order. Most all radio polarization maps are 
much more complicated, with hundreds of examples not looking like 
this one! So what kind of selection is going on? Not only does the 
map prove nothing -- nobody knows, remember, what happened back at 
the source -- but it reveals a highly unscientific method of throwing 
away ugly objects and basing conclusions on pretty objects.  

An underlying persistent problem is the implicit assumption that ``one
knows for sure'' the emission angle of polarization at the galaxy
source. Now, if synchrotron radiation is producing the polarization (a
pretty good model), then the polarization at the emission point will be
perpendicular to the magnetic field at the same point. But the facts
are that nobody knows the magnetic fields at the source, and nobody
knows the polarization angle at the source, as neither has ever been
measured. If one would argue that the data, by definition, shows us the
same observed polarization at the antenna on Earth as that emitted at
the source, then one is engaging in a circular rebuttal, which cannot
even permit the question of cosmological birefringence to be asked.

Such false reasoning seems to us to be the nature of Leahy's
claims, because, while he says that ``the projected magnetic field is
predominantly perpendicular to strong gradients ... in excellent
agreement with theory,'' the evidence for agreement is entirely
circular. There is no evidence aside from the inferences based on
models built to agree with the interpretation Leahy chooses to use.
Nor has Leahy provided a convincing statistical basis for the assertion
itself, which, as we said above, is based on selecting the parts of
pictures that look pretty. History shows that the first observed
orientations of polarization at high resolution were a surprise, and
just about at right angles to the polarization orientation deduced from
dimensional analysis and scaling arguments. This happens to be almost
what the de-rotated polarization orientation would give for most of the
cosmologically distant ($z = 1$) samples. 

In addition, the magnetic field is not thought to be oriented in a
single direction, but to vary over the source. A further complication
is whether one knows which region is bright enough to dominate in
synchrotron radiation: the brightest radiating parts are not
necessarily the ones with the nicest fields. It does seem that VLBI
data could be used productively in a responsible statistical way. But
we seriously question the generality of conclusions circularly
supported by empirical observations in the previous framework. The glib
suppression of numerous complications known to experts like Leahy,
which seems to us to be a form of unscientific cover--up of the facts,
is deplorable.

We observed something very simple: an anisotropic correlation between
observed polarization orientations and observed galaxy orientations. We
reported what we found. That's it. Even with the shuffling analysis
Eisenstein and Bunn suggested (with which we don't agree), the signal
of anisotropy persists \cite{nodeis}. Anyone can dig through sources to
find particular cases, without proving a thing. Until the physics is
understood, it seems irresponsible and unscientific to dismiss our
correlation using different variables, inventing incompatible measures,
ignoring the need for statistics, egregiously selecting data, and
misreporting and misrepresenting what we found. On the theoretical
side, constructive work is carried out \cite{obu,dob,bra}. What is needed,
to make further progress on the experimental side, is an order of
magnitude or two orders of magnitude more data. The subject needs
10,000 sources. With 10,000 sources, tremendous positive science could
be done to pinpoint the origin of the anisotropy we observed. Why those
interested in radio astronomy would so strenuously reject this
challenge is the most incredible comment yet made on the subject.

\end{document}